\documentclass[12pt,letterpaper]{article}
\usepackage{jheppub,amsmath}
\usepackage{natbib}
\usepackage{graphicx}
\usepackage{caption}
\usepackage{subcaption}

\makeatletter
\def\@fpheader{\relax}
\makeatother
\subheader{}
\title{Unitarity From a Smooth Horizon?}

\author[a,b]{Raphael Bousso,}
\affiliation[a]{Center for Theoretical Physics and Department of Physics,\\
  University of California, Berkeley, CA 94720, U.S.A.}
\affiliation[b]{Lawrence Berkeley National Laboratory, Berkeley, CA 94720,
   U.S.A.}
\author[c]{and Marija Toma\v{s}evi\'{c}}
 \affiliation[c]{Departament de F\'isica Qu\`antica i Astrof\'isica and Institut de Ci\`encies del Cosmos (ICCUB),\\
  Universitat de Barcelona, Mart\'i i Franqu\`es 1, E-08028 Barcelona, Spain}

\abstract{Under semiclassical evolution, black holes retain a smooth horizon but fail to return information. Yet, the Ryu-Takayanagi prescription computes the boundary entropy expected from unitary CFT evolution. We demonstrate this in a novel setting with an asymptotic bulk detector, eliminating an assumption about the entanglement wedge of auxiliary systems.
  
  We consider three interpretations of this result. (i) At face value, information is lost in the bulk but not in the CFT. This conflicts with the AdS/CFT dictionary. (ii) No unique QFT state (pure or mixed) governs all detector responses to the bulk Hawking radiation. This conflicts with the existence of an S-matrix. (iii) Nonlocal couplings to the black hole interior cause asymptotic detectors to respond as though the radiation was pure, even though it is naively thermal. This invalidates the standard interpretation of the semiclassical state, including its smoothness at the horizon.

  We conclude that unitary boundary evolution requires asymptotic bulk detectors to become unambiguously pure at late times. We ask whether the RT prescription can still reproduce the boundary entropy in this bulk scenario. We find that this requires a substantial failure of semiclassical gravity in a low-curvature region, such as a firewall that purifies the Hawking radiation.

  Finally, we allow that the dual to semiclassical gravity may be an ensemble of unitary theories. This appears to relax the tensions we found: the ensemble average of out-states would be mixed, but the ensemble average of final entropies would vanish.}

\begin{document}
\maketitle

\section{Introduction}

\subsection{Scattering and the Information Paradox}

The information paradox was first formulated for black holes in asymptotically flat spacetime. The S-matrix is expected to be unitary, so pure in-states should be mapped to pure out-states. The S-matrix is an asymptotic observable even in the presence of gravity, since gravity becomes weak in a dilute out-state. But Hawking showed that a black hole evaporates into an approximately thermal Hawking cloud, regardless of how it was formed~\cite{Haw74,Haw76}.

Hawking's result followed from a semiclassical calculation: one solves
\begin{equation}
  G_{\mu\nu} = 8\pi G \langle T_{\mu\nu}\rangle
\end{equation}
iteratively in powers of $G\hbar$. Here $\langle T_{\mu\nu}\rangle= {\rm Tr} (\rho T_{\mu\nu})$, where $\rho=\rho_{\rm Hawking}$ is the global state of the quantum fields. This state is pure at all times. Information is lost because the asymptotic observer has no access to the black hole interior. Tracing over the interior gives the mixed out-state:
\begin{equation}
  \rho_{\rm out,Hawking} = {\rm Tr}_{\rm in}\, \rho_{\rm Hawking}~.
\end{equation}

The semiclassical approximation should receive nonperturbative corrections, and these may restore the unitarity of the S-matrix. But this comes a steep price. If effective field theory is valid outside the horizon, a pure out-state implies that a freely falling observer encounters large excitations (a ``firewall'') at the horizon of an arbitrarily large black hole, at least after the Page time~\cite{AMPS,AMPSS}. (The Page time $t_{\rm Page}$ is when the coarse-grained entropy of the radiation first exceeds the Bekenstein-Hawking entropy of the black hole.)

An interesting class of approaches~\cite{PapRaj12,MalSus13,PapRaj13b} constructs effective interior operators consistent with a smooth horizon. But this works only for certain classes of states, and only at the cost of introducing significant nonlinearity in the form of state-dependence~\cite{Bou12c,Bou13,MarPol13,Bou13a}. It remains to be seen whether these ideas can be developed into a consistent framework that preserves both unitarity and the equivalence principle. (See Refs.~\cite{Bou13b,MarPol15,Har14a} for some challenges; see Ref.~\cite{Har14b} for a review and further references.)

\subsection{Quantum Ryu-Takayanagi Prescription and Recent Work}

The AdS/CFT correspondence~\cite{Mal97} constitutes the most significant evidence that the S-matrix remains unitary in the presence of gravity. The initial and final states of a bulk (AdS) scattering experiment can be mapped to states in the CFT. The CFT is manifestly unitary, so these bulk states are related by a unitary operator.

However, this does not explain how the information comes out from a bulk perspective. AdS/CFT has not told us whether and how firewalls form, or if not, how they are evaded. Recent works by Penington~\cite{Pen19} and by Almheiri {\em et al.}~\cite{AEMM} have the potential to shed some light on this question. Let us briefly review some background.

The generalized entropy $S_{\rm gen}$~\cite{Bek72} of a surface $\sigma$ is the sum of its area and the von Neumann entropy of the quantum fields in its exterior:
\begin{equation}
  S_{\rm gen}[\sigma] =\frac{{\cal A}(\sigma)}{4G\hbar}+S[\mbox{Ext}(\sigma)]~.
\end{equation}
A Quantum Extremal Surface (QES) is a surface whose generalized entropy is stationary with respect to all deformations. Such surfaces play a central role in the quantum-corrected~\cite{FauLew13,EngWal14} Ryu-Takayanagi~\cite{RyuTak06}/Hubeny-Rangamani-Takayanagi~\cite{ HubRan07} prescription, which we now briefly summarize.

The von Neumann entropy of a holographic CFT restricted to a given boundary region $R$ can be computed from the bulk dual as
\begin{equation}
  S_{\rm CFT}[R] = S_{\rm gen}[\mbox{Ext}(\gamma_{\rm min}[R])]~.
  \label{eq-rt}
\end{equation}
Here $\gamma_{\rm min}$ is the QES with {\em smallest} generalized entropy homologous to $R$; and Ext$(\gamma_{\rm min})$ is chosen to be the bulk region bounded by $R\cup \gamma_{\rm min}$. This region is called the entanglement wedge of $R$ and will be denoted $EW(R)$. 

\begin{figure}
\centering
\begin{subfigure}{.5\textwidth}
  \centering
  \includegraphics[width=.8\linewidth]{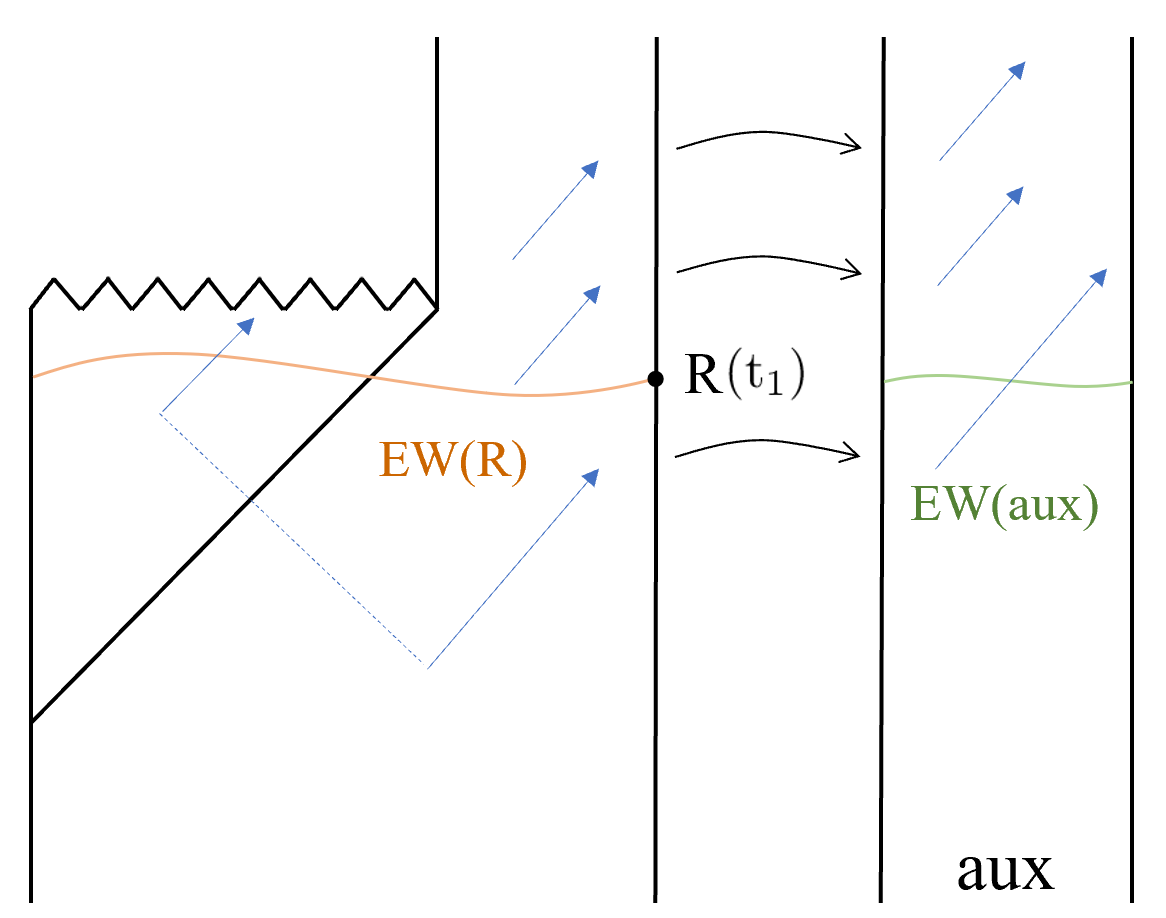}
  \caption{ }
  \label{fig:sub11}
\end{subfigure}%
\begin{subfigure}{.5\textwidth}
  \centering
  \includegraphics[width=.9\linewidth]{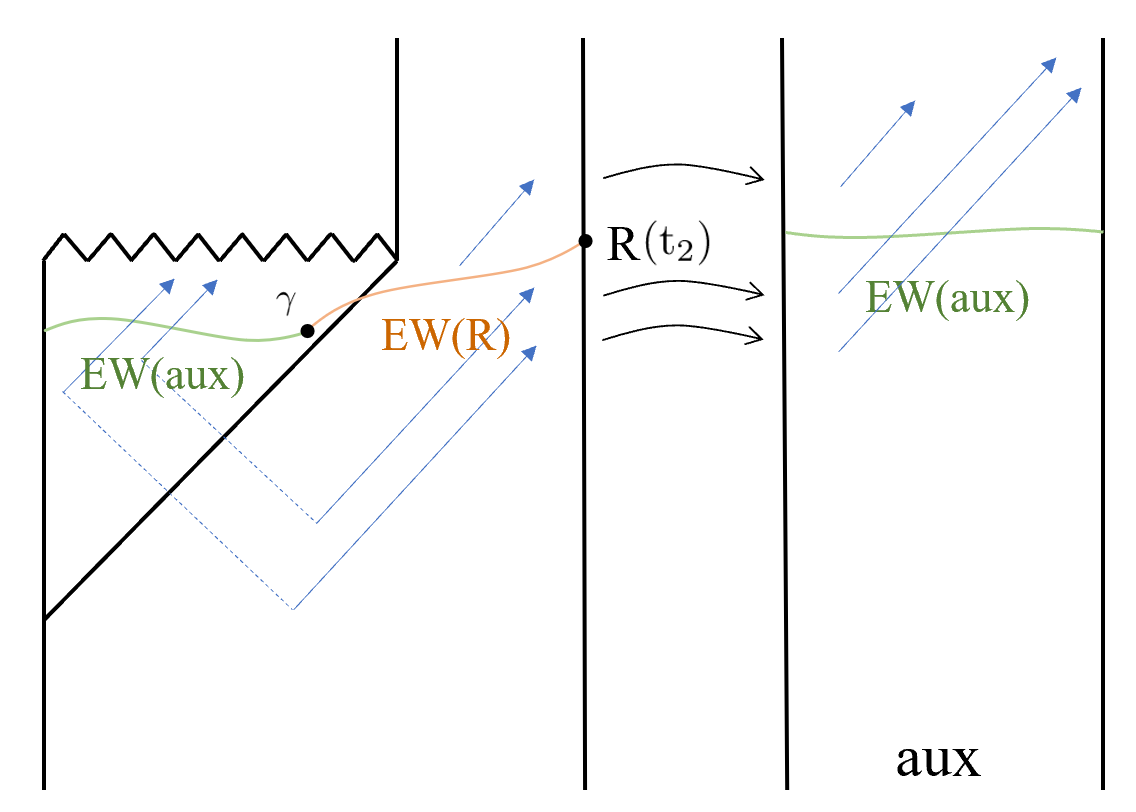}
  \caption{ }
  \label{fig:sub22}
\end{subfigure}
\caption{Semiclassical bulk evolution of a black hole in AdS with global boundary $R$. The Hawking radiation is absorbed into an auxiliary system~\cite{Pen19,AEMM}. The entanglement wedges $EW(R)$ and $EW({\rm aux})$ are shown (a) before and (b) after the Page time. Entanglement wedge complementarity is assumed here but will not be needed in the setting we describe in Sec.~\ref{sec-dyson}.}
\label{fig-qes}
\end{figure}

Refs.~\cite{Pen19,AEMM} applied the RT prescription in a peculiar setting. (See Refs.~\cite{AMMZ,AkeEng19,AlmMah19,RozSul19,CheFis19} for some discussions and extensions.) The bulk evolution is computed semiclassically, using the state $\rho_{\rm Hawking}$. In this description, the horizon is manifestly smooth. The Hawking radiation is allowed to escape from the AdS spacetime into an external bath. Choosing $R$ to be the entire boundary of the original AdS spacetime containing the black hole, Refs.~\cite{Pen19,AEMM} discovered a novel QES (Fig.~\ref{fig:sub22}): $\gamma(t)$ is located approximately one Planck length inside the horizon, at about one scrambling time before $t$:
\begin{equation}
  \Delta t_s\sim \beta \log (S-S_0)~.
\end{equation}
Here $\beta$ is the inverse Hawking temperature, $S$ is the Bekenstein-Hawking entropy of the black hole, and $S_0$ is the ground state entropy (for charged black holes).

The newly discovered QES $\gamma(t)$ competes with the trivial QES, $\varnothing$. (The empty surface satisfies the homology constraint, since the boundary sphere can be contracted to a point; and it is stationary since there are no points to deform.) Ext$(\varnothing)$ comprises the entire original bulk, whereas Ext$(\gamma)$ consists only of the horizon and black hole exterior.\footnote{The above discussion pertains to a one-sided black hole formed from collapse~\cite{Pen19}. For a two-sided (eternal) black hole~\cite{AEMM}, one may choose $R$ to be the union of the right and left boundary CFT. Then the newly discovered QES $\gamma$ has two components, near the left and right black hole horizon. One could also consider a single component of the boundary. In this case, the new QES competes with the bifurcation surface $\gamma_0$.}

One finds that before the Page time,\footnote{If matter is added to the black hole, then the Page transition can occur at multiple times. A new QES of the type discovered in Refs.~\cite{Pen19,AEMM} will form on every such occasion as soon as the horizon settles down.} $\varnothing$ is the minimal QES. Hence there is no area term and the RT prescription yields $S[R]= S_{\rm bulk}$, where $S_{\rm bulk}$ is the global von Neumann entropy in the bulk. In the semiclassical analysis, the black hole interior exactly purifies the Hawking radiation, so their von Neumann entropies are equal. Since the radiation is moved to an external system, the bulk von Neumann entropy is that of the interior ``Hawking partner modes.'' Hence
\begin{equation}
  S[R](t) = S_{\rm rad}(t)~,~~(t<t_{\rm Page})~,
\end{equation}
where $S_{\rm rad}$ is the entropy of the Hawking radiation that has been emitted and transferred to the auxiliary system by the time $t$. This quantity grows monotonocially.

After the Page time $t_{\rm Page}$, $\gamma$ becomes the minimal QES, because then $S_{\rm gen}(\gamma(t))=A/4G\hbar<S_{\rm rad}$ by definition of the Page time. Hence
\begin{equation}
S[R](t) = \frac{A(t)}{4G\hbar}~,~~(t>t_{\rm Page})~,
\end{equation}
where $A$ is the area of the black hole. This quantity decreases monotonically.

Therefore, the entropy $S_{\rm CFT}[R]$ follows a Page curve: the entropy grows from 0 to a maximum at the Page time, so long as $\gamma_{\rm min}=\varnothing$. Then it shrinks back to 0, while $\gamma_{\rm min}=\gamma$. This is exactly as expected from unitary evolution of the CFT. But it is interesting that it is reproduced by applying the RT prescription to the semiclassically evolved bulk---precisely the type of evolution that leads to information loss for asymptotic observers.% (Note that without the new QES, the boundary computed by RT would have increased at all times.)

The result becomes even more puzzling when we consider the auxiliary system, which contains the Hawking radiation. The bulk calculation says that this radiation is mixed. But on the other hand, suppose we choose the auxiliary system to be another CFT (perhaps with much larger central charge), with its own bulk. One could speculate that {\em its} entanglement wedge, $EW({\rm aux})$, should be the complement of $EW(R)$. Under this assumption $EW(\rm aux)$ should include the {\em interior} of the QES $\gamma(t)$ after the Page time. After the black hole has disappeared, $EW(\rm aux)$ would still include the black hole interior as a disconnected universe. In particular this would mean that local operators in the interior can be realized as operators with support on ${\rm aux}$ and hence, presumably, as operators on the Hawking radiation.

To summarize, the results of Refs.~\cite{Pen19,AEMM} are intriguing and puzzling. Bulk evolution is computed semiclassically, which should result in information loss; % to a bulk observer. That is, the Hawking radiation is in a mixed state, not a pure state. Yet
yet the RT prescription ``fails to fail.'' It predicts a boundary entropy consistent with unitarity, from a bulk calculation that is not.

\subsection{Outline and Summary}

Our analysis of Refs.~\cite{Pen19,AEMM} is motivated by the original information problem, as posed in an asymptotically flat region. We are interested in what happens in a (futuristic) real-world experiment where a black hole is formed in a laboratory and is allowed to evaporate. Is information returned to the laboratory? Is the horizon smooth after the Page time?

We will assume that an analysis in AdS using the RT prescription must reproduce features that are essential from this operational viewpoint. Otherwise, the AdS analysis would have no relevance to the actual information problem. In particular, we will insist that the response of detectors in distant regions, in experiments which engender no large gravitational backreaction, is fully described by quantum field theory, to arbitrarily good approximation.

This is an assumption, of course; but it would be quite interesting if it was false. It would mean that the validity of QFT, which has been confirmed through numerous experiments, is under no clear control in any regions of spacetime, including weakly gravitating asymptotic regions. The very notion of an S-matrix would be in question.

In Sec.~\ref{sec-dyson} we reproduce the key results of Refs.~\cite{Pen19,AEMM} in a setting that is closely analogous to a real-world scattering experiment.

In order to render the AdS setting as similar as possible to the laboratory setting, we introduce a large detector sphere (Dyson sphere), in Sec.~\ref{sec-dysondetectors}. This sphere lives in the bulk of AdS, but far from the black hole. Because of its arbitrarily large size, it can absorb all of the Hawking radiation, and complicated experiments can be performed without large backreaction.

Our setting has no auxiliary system; thus it requires no assumptions about the entanglement wedge of such a system. It also leaves no ambiguity as to what degrees of freedom correspond to the Hawking radiation in a real-world experiment. (In Ref.~\cite{Pen19}, there is both a bulk and boundary auxiliary system, leaving some ambiguity on this point.)

We perform an RT analysis analogous to that of Refs.~\cite{Pen19,AEMM} in our setting. We allow the bulk to evolve semiclassically (Sec.~\ref{sec-hawking}). We then apply the RT prescription to compute the entropy of the boundary CFT. In Sec.~\ref{sec-success}, we consider the entire boundary. Since there is no auxiliary, the boundary should remain in a pure state at all times. We find that the bulk analysis reproduces this, for the simple reason that there is no nontrivial QES at any time, and that the global bulk state $\rho_{\rm Hawking}$ (including the black hole interior) is pure.

In Sec.~\ref{sec-fancy}, we refine our setup by transferring the absorbed Hawking radiation to an angle-localized reservoir on the Dyson sphere.  This allows us to compute separately the entropy of a boundary region dual to just the Hawking radiation, and the entropy of a complementary boundary region dual to the complementary bulk region that includes the black hole interior. We find that both follow the Page curve. The novel QES of Refs.~\cite{Pen19,AEMM} now makes a crucial appearance. But unlike in Refs.~\cite{Pen19,AEMM}, entanglement wedge complementarity is enforced by the usual homology constraint and need not be assumed.

In Sec.~\ref{sec-interpretations}, we try to make sense of the apparent contradiction inherent in these results: information is lost in the bulk but not in the CFT. The Dyson sphere is in a mixed state after the black hole is gone, but boundary evolution is unitary\footnote{It may be possible to interpret the semiclassical bulk calculation in terms of an ensemble of boundary theories~\cite{SaaShe18,SaaShe19}. We consider this possibility in the discussion.} and the RT calculation confirms this.

In Sec.~\ref{sec-infoloss}, we take this outcome at face value. For a real-world experiment, this interpretation would imply information loss. The presence of the information in some inaccessible boundary theory would be irrelevant. However, we argue that this interpretation conflicts with the AdS/CFT dictionary. For energetic reasons, the boundary state dual to $\rho_{\rm Hawking}$ cannot be pure.\footnote{For the doubly-holographic analysis of Ref.~\cite{AMMZ,AlmMah19}, this result implies that the semiclassically evolved JT-brane does not have a pure-state lower-dimensional BCFT dual (though it can have a higher-dimensional bulk dual). This is relevant for determining which homology constraint~\cite{RyuTak06,Tak11} applies.}

In Sec.~\ref{sec-firstcommandment}, we allow for the possibility that no single QFT state (pure or mixed) governs all detector responses to the bulk Hawking radiation. In this interpretation, $\rho_{\rm Hawking}$ should be used for evaluating the experience of an infalling observer, and it could optionally be used to predict the results of simple probes of the Hawking radiation. By contrast, sufficiently complicated experiments would reveal the unitarity of the scattering problem to an asymptotic bulk observer. They would therefore have to be described by a different state $|\Psi\rangle_{\rm Dyson}(t_{\rm late})$, in which the Dyson sphere becomes pure at late times. However, we find that this possibility conflicts with our assumption that quantum field theory is valid in asymptotic, weakly gravitating regions. There would be no S-matrix for asymptotic observers.

In Sec.~\ref{sec-dual}, we consider the possibility that all asymptotic detectors respond to the pure out-state $|\Psi\rangle_{\rm Dyson}(t_{\rm late})$ according to the standard rules of local QFT. This description, however, is to be viewed as a kind of {\em effective} theory that results from tracing over the black hole interior. The same detector responses can also be predicted from $\rho_{\rm Hawking}$, viewed here as a more fundamental description, by invoking nonlocal couplings of the asymptotic detectors to the black hole interior. This would imply that detectors in the $\rho_{\rm Hawking}$-description do not respond as required by local QFT. But in the state $\rho_{\rm Hawking}$, detectors would see a smooth horizon if they {\em did} respond as required by local QFT. Hence, the naive smoothness of $\rho_{\rm Hawking}$ would not guarantee the absence of drama for an infalling observer.

In Sec.~\ref{sec-unitary}, we discuss the contrapositive of our conclusions in Sec.~\ref{sec-dual}. Since $\rho_{\rm Hawking}$ cannot be dual to a pure boundary state by the extrapolate dictionary, and assuming that the state of the asymptotic region is an unambiguous QFT state, the Dyson sphere must end up in the pure state $|\Psi\rangle_{\rm Dyson}(t_{\rm late})$. We ask whether there exists a global bulk state consistent with this restriction, and with the property that the RT prescription gives the correct boundary entropy. We find that this is impossible unless effective field theory breaks down in a low-curvature region in the bulk. However, it is not necessary to invoke such a breakdown in the asymptotic region. If the early Hawking radiation is purified by a physical structure at the horizon (a firewall), then effective field theory and black hole thermodynamics can at least be preserved in the exterior.

In Sec.~\ref{sec-discussion}, we discuss the implications of our results for the information paradox, and we suggest a possible interpretation in terms of an ensemble of boundary theories.

\section{Boundary Entropy From Semiclassical Bulk Evolution}
\label{sec-dyson}

With reflecting boundary conditions, sufficiently large black holes in Anti-de Sitter space will not evaporate, so the question of information loss cannot be posed operationally as a scattering problem. Evaporation can be implemented by imposing absorbing boundary conditions, whereby the radiation is transferred to an auxiliary system. This approach was recently taken in Ref.~\cite{Pen19}, and for a two-sided black hole in Ref.~\cite{AEMM}, who computed the entropy of the boundary theory and of the auxiliary system using the Ryu-Takayanagi (RT) proposal~\cite{RyuTak06,HubRan07,FauLew13,EngWal14}.

However, the auxiliary system does not live in the same spacetime as the black hole. We would like to avoid any ambiguities or complications that such a setup may lead to, while still using the RT proposal to compute the entropy of the boundary theory. In particular, the entanglement wedge of the auxiliary system is ambiguous unless one assumes entanglement wedge complementarity~\cite{AEMM}. Here we will be able to justify this choice.

\subsection{Large Dyson Sphere as a Detector in AdS}
\label{sec-dysondetectors}

Indeed, there are alternative ways of allowing a black hole to fully evaporate in AdS. One possibility is to consider small enough black holes, with $t_{\rm evap}<L$, where $L$ is the AdS length. However, this restriction is not necessary if we include a detector sphere with large radius $d\gg L$ (i.e., ``near infinity''). We will refer to this as a Dyson sphere.

The Dyson sphere can be viewed as a laboratory in which the entire scattering experiment takes place: it prepares the in-state and it measures the out-state. The Hawking radiation is absorbed into a reservoir located on the Dyson sphere. Before the first particle comes out, the reservoir is initialized in a fiducial state $|0\rangle_{\rm Dyson}$.  At any later time, an observer on the Dyson sphere may choose to probe the state of all or parts of the Hawking radiation.

Here we assume that there exists a description of the out-state on the Dyson sphere as a quantum field theory state (in the sense of QFT on a fixed background). This description becomes exact in the large radius limit. We will discuss this assumption in more detail in Sec.~\ref{sec-firstcommandment}.

The mass and complexity of the Dyson sphere is not limited by fundamental considerations such as entropy bounds. Its area will be exponential in its proper radius. Therefore, one can consider the evaporation of an arbitrarily large black hole in AdS. In this regime, $t_{\rm evap}\gg d\gg L$. We will not distinguish between large and small black holes in what follows.

%V2
A Dyson sphere in AdS must be stabilized against the gravitational potential, e.g., with rods or by giving it an intrinsic tension, like a brane. A static Dyson sphere in AdS can have entropy proportional to its area~\cite{BouFre10a}, a remarkable property not shared by spheres in asymptotically flat spacetime. (We thank B.~Freivogel for reminding us of this result.) However, we are not aware of in-principle obstructions. Moreover, if the assumption of a Dyson sphere failed, this would mean that the information problem cannot be operationally posed for large AdS black holes by observers in AdS. If so, their study would not allow for reliable conclusions about experiments that can actually be carried out, such as the formation and evaporation of black holes in asymptotically flat spacetime.

\subsection{Semiclassical Evaporation in the Bulk}
\label{sec-hawking}

We now consider the formation and evaporation of an AdS black hole in the presence of a Dyson sphere. As described above, the sphere is initialized in the reference state $|0\rangle_{\rm Dyson}$ as the black hole forms. It then absorbs all of the radiation.

Inspired by Ref.~\cite{Pen19,AEMM}, we will describe the bulk evolution by Hawking's semiclassical analysis~\cite{Haw74}. That is, we compute the out-state using QFT on a curved Schwarzschild background.
\begin{figure}%
    \centering
    \includegraphics[width=.5\textwidth]{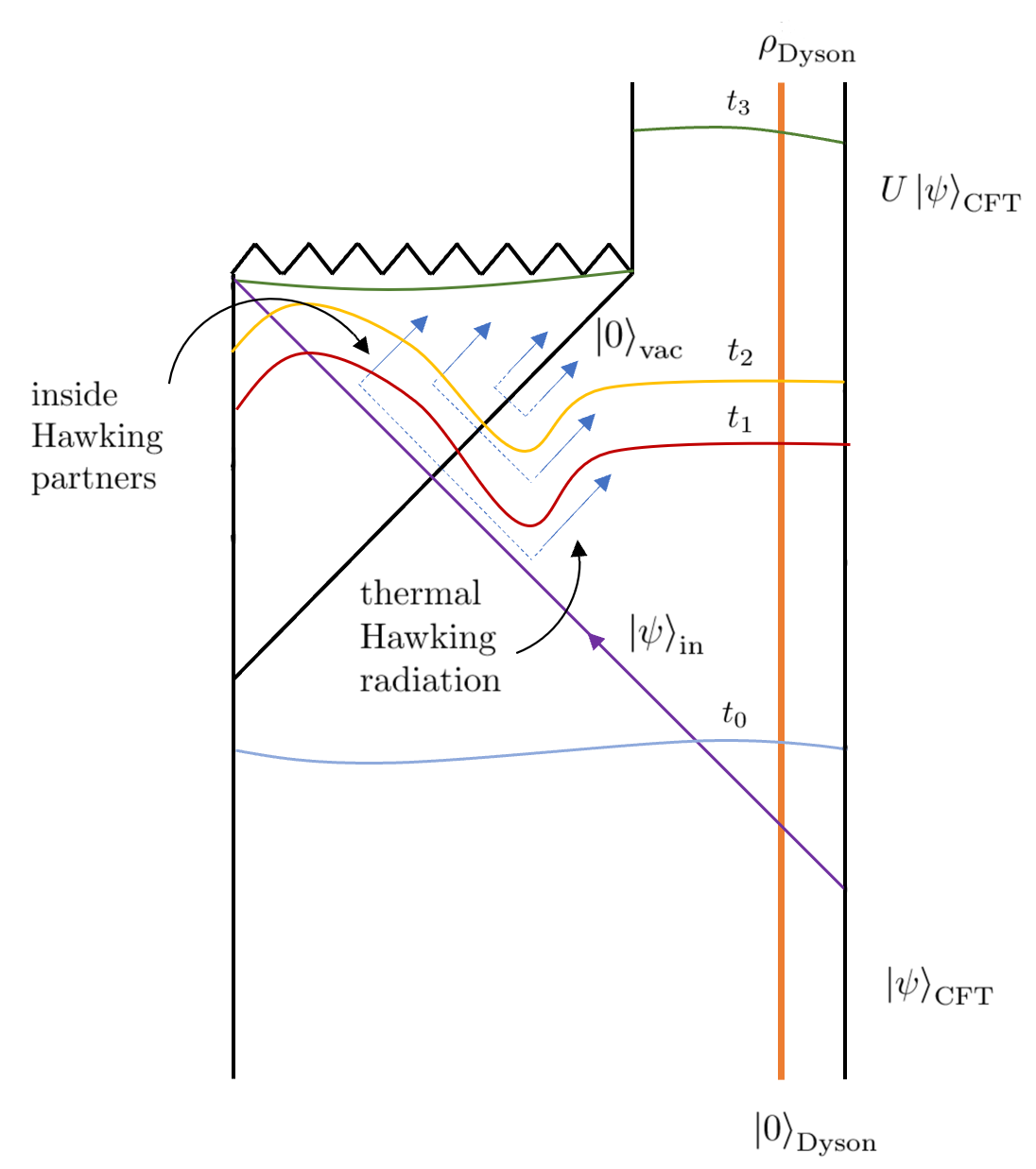}
    \caption{Formation and evaporation of a black hole in AdS. The Hawking radiation is absorbed into a Dyson sphere near the boundary. The bulk evolution is computed semiclassically. Nevertheless, the Ryu-Takayanagi prescription yields a boundary entropy consistent with unitary boundary evolution. However, energetic arguments and the extrapolate dictionary imply that the semiclassical bulk state at late times cannot have a pure-state boundary dual (see Sec.~\ref{sec-interpretations}). This conclusion depends only on the largeness of the entropy of the Hawking radiation in the bulk. Because the Dyson sphere can be probed with arbitrarily dilute local operators, even complicated bulk probes of the Dyson sphere do not engender large gravitational backreaction, and standard QFT rules should apply.}%
    \label{fig-dyson}%
\end{figure}

In this picture, the {\em global}\/ state in the bulk, $\rho_{\rm Hawking}$ is always pure (Fig.~\ref{fig-dyson}). Initially, it consists of the Dyson sphere and the collapsing matter, each in a pure state:
\begin{equation}
  \rho_{\rm Hawking}(t_0) = |\psi\rangle_{\rm in}\mbox{}_{\rm in}\langle\psi| \otimes |0\rangle_{\rm Dyson}\mbox{}_{\rm Dyson}\langle 0|~.
\end{equation}
After the black hole has formed, the bulk can be thought of as consisting of three subsystems. The first is the collapsed matter inside the black hole, in the state $|\psi\rangle_{\rm in}\mbox{}_{\rm in} \langle\psi|$. The second is the (mixed) interior subsystem of the (pure) vacuum state spanning the horizon. The third is the (mixed) exterior subsystem of the vacuum, which becomes the Hawking radiation and which is absorbed into the Dyson sphere. Schematically,
\begin{equation}
  |0\rangle_{\rm vacuum} = N \prod_{\omega} \sum_{n=0}^\infty e^{-\beta n\omega/2} |n\rangle_{\rm inside}\otimes  |n\rangle_{\rm outside}~,
\end{equation}
where $\beta$ is of order the black hole radius, and $\omega$ labels modes with support strictly inside or outside the horizon.

The von Neumann entropy of the Dyson sphere grows as it absorbs the thermal radiation. At the same time, the von Neumann entropy of the black hole interior grows due to the accumulation of inside partners of the outgoing Hawking radiation. These two systems purify each other at all times. Their individual entropy increases strictly monotonically, until the black hole has fully evaporated. Neither system obeys a Page curve.

All bulk probes of the Dyson sphere are fully described by the state of the Dyson sphere, which is mixed due to the absorption of thermal Hawking radiation in this model. Therefore, information is lost to a bulk observer; probes of the Dyson sphere would not be able to reconstruct the pure state from which the black hole was formed. 

\subsection{Boundary Unitarity From the RT Prescription}
\label{sec-success}

An important ingredient in the AdS/CFT dictionary is that the entropy of a boundary region equals the generalized entropy of the entanglement wedge in the bulk, i.e., the area of the associated RT surface~\cite{RyuTak06} plus the entropy of the bulk matter in the enclosed region~\cite{FauLew13,EngWal14}:
\begin{equation}
  S_{\rm CFT} = \frac{A_{RT}}{4G\hbar}+ S_{\rm bulk}~.
  \label{eq-flm}
\end{equation}
We will now verify this relation in our example.

The boundary state is pure initially. It remains pure by unitarity of the CFT, so
\begin{equation}
  S_{\rm CFT}=0
\end{equation}
at all times. But the bulk state is computed only semiclassically, and this leads to information loss in the bulk. Thus, one might naively expect that Eq.~(\ref{eq-flm}) will fail. 

However, the RT surface associated with the entire boundary is always the trivial (or empty) surface. That is, the entanglement wedge includes the entire bulk at all times. And as we have noted, the global bulk state is indeed pure. Hence
\begin{equation}
  A_{RT}=0~,~~S_{\rm bulk}=0
\end{equation}
at all times, and Eq.~(\ref{eq-flm}) holds.

This analysis is different from, and simpler than, the case where radiation is extracted from the bulk~\cite{Pen19,AEMM}. (Indeed, our main motivation in including a Dyson sphere was to allow us to consider this simple scenario where no extraction is needed.) In our setup, the quantum extremal surface near the horizon never dominates in the RT prescription, since the exterior radiation is not removed. The radiation is merely absorbed into the Dyson sphere, so it remains in the bulk.

We stress that this agreement comes about {\em not} because the bulk Hawking radiation is pure in this model. The entanglement wedge of the whole boundary includes the black hole interior. This is obvious both before and after the Page time ($t_1$ and $t_2$ in Fig.~\ref{fig-dyson}), when the black hole has not fully evaporated. Continuity at the endpoint of evaporation makes it natural at $t_3$ to include the pinched-off black hole interior in the entanglement wedge, which then again leads to agreement with Eq.~(\ref{eq-flm}).

Thus our single-bulk example shares the feature~\cite{Pen19,AEMM} that the boundary entropy expected from unitary boundary evolution is correctly reproduced by applying the RT prescription to a semiclassically evolved bulk. In Refs.~\cite{Pen19}, the boundary information was distributed over two systems. Unitarity required that they obey the Page curve, and they were found to do so using RT. However, this required an additional assumption. We next consider a bipartite version of our setup in which the Page curve is recovered with no additional assumptions.

\subsection{Boundary Page Curve from a Bulk Island}
\label{sec-fancy}

In this subsection we consider a refinement of the previous setup, more closely analogous to the bipartite configurations studied in Refs.~\cite{Pen19,AEMM}. Consistent with these works, we will show that the RT prescription applied to a semiclassically evolved bulk reproduces the Page curve for the boundary dual of each relevant subsystem: the dual to the black hole, and the dual to the Hawking radiation.

However, in those works an ambiguity was encountered (as stressed in~\cite{AEMM}): in order to get the answer demanded by unitarity, one had to assume that the bulk dual of the auxiliary system outside of the original spacetime should be the complement of the dual of the original CFT, and so should include the black hole interior after the Page time. This has been criticized~\cite{AEMM,AMMZ} as tantamount to putting in the desired answer.

In our analysis below, we will need not to assume this. We have only a single boundary, and the inclusion of the interior will follow from the usual homology condition in the RT proposal. 
\begin{figure}
\centering
\begin{subfigure}{.5\textwidth}
  \centering
  \includegraphics[width=.8\linewidth]{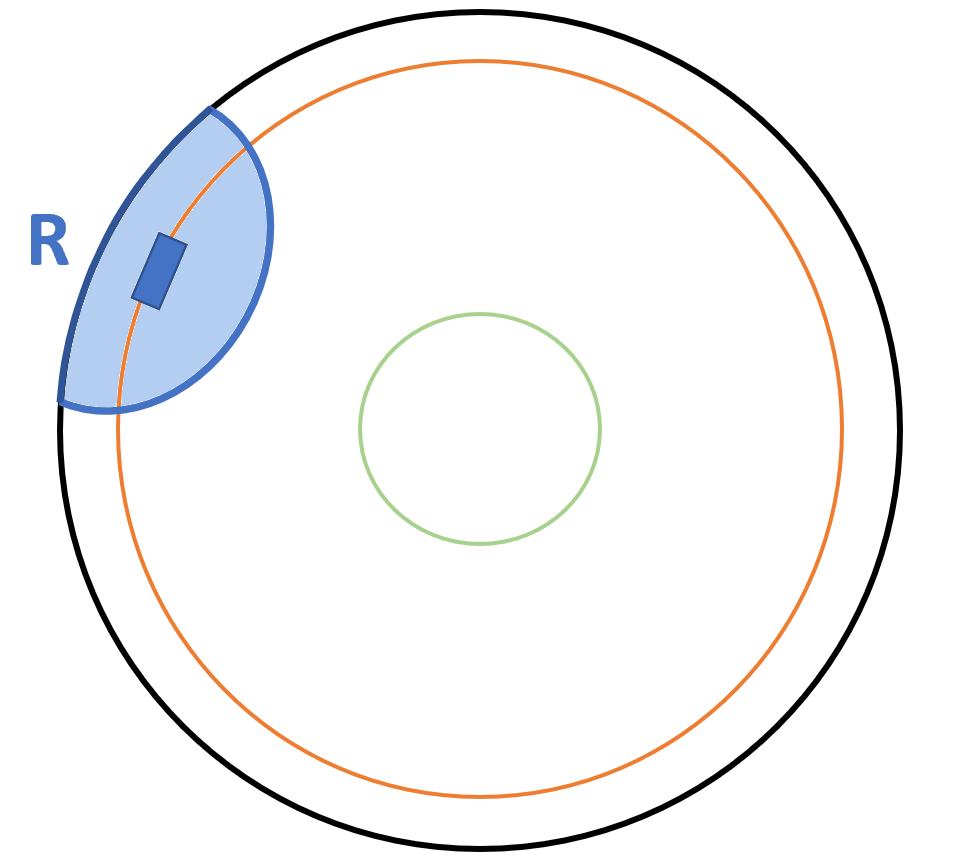}
  \caption{ }
  \label{fig:sub1}
\end{subfigure}%
\begin{subfigure}{.5\textwidth}
  \centering
  \includegraphics[width=.8\linewidth]{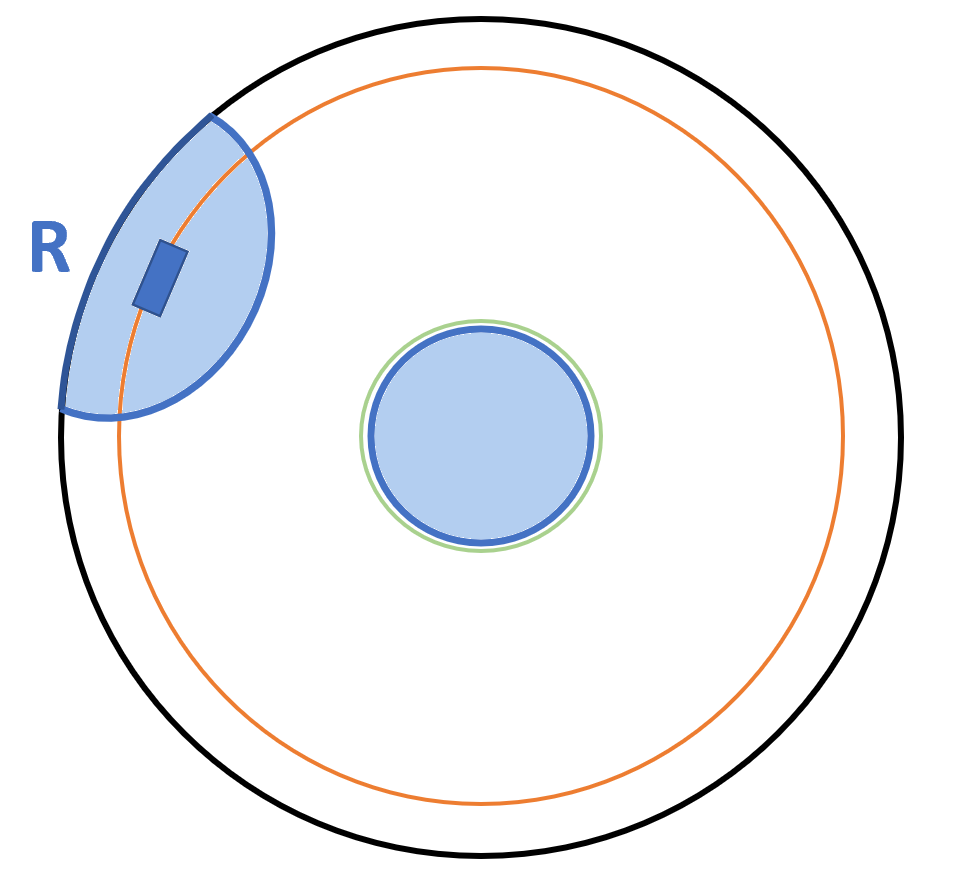}
  \caption{ }
  \label{fig:sub2}
\end{subfigure}
\caption{In a semiclassically evolved bulk state, the Hawking radiation is absorbed and transferred to a near-boundary reservoir, localized to a small angle. $R$ is a boundary region near the reservoir. (a) At $t_1<t_{\rm Page}$, the entanglement wedge $EW(R)$ includes only the reservoir. (b) At $t_2>t_{\rm Page}$, the minimal quantum extremal surface $\gamma$ has a second component near the black hole horizon. $EW(R)$ now contains the black hole interior.}
\label{fig-localdyson}
\end{figure}

We use the same setup as before. But now we localize the reservoir to a particular region of small angular scale $\delta_{\rm res}$ (but arbitrarily large physical scale) on the Dyson sphere; see Fig.~\ref{fig-localdyson}.  The radiation is absorbed at all angles, but then it is transferred coherently along quantum channels in the Dyson sphere, into the reservoir. 

Let $R$ be a connected, ball-shaped boundary region centered on the angular position of the reservoir, with angular radius $\delta_R$. We choose
\begin{equation}
  \beta\gg \delta_R\gg \delta_{\rm res}~,
\end{equation}
where $\beta$ is the characteristic boundary wavelength associated to the black hole. With this choice, the entanglement wedge of R will include the reservoir at all times and yet its component connected to $R$ will stay far from the black hole. The complement region on the boundary is denoted by $\bar R$.
%Let $\bar R$ be the complement region on the boundary.

We now apply the quantum-corrected RT prescription~\cite{FauLew13,EngWal14} to compute $S_R(t)$ and $S_{\bar R}(t)$, as the generalized entropy of $EW(R)$ and $EW(\bar R)$.

The semiclassically evolved global bulk state is pure at all times, so the bulk entropies on two sides of any surface must agree. This implies entanglement wedge complementarity in this setting. That is, $R$ and $\bar R$ will have the same minimal-$S_{\rm gen}$ quantum extremal surface $\gamma(t)$. Its complementary exteriors define the respective entanglement wedges $EW(R)$, $EW(\bar R)$, which will have the same $S_{\rm gen}$. Therefore,
\begin{equation}
  S_R(t)=S_{\bar R}(t)
\end{equation}
at all times. This is consistent with unitary evolution of the pure boundary state. We stress that in our setting this is an implication of RT, not an assumption.

Both entropies contain a divergent piece from vacuum entanglement around $\partial R$ on the boundary. In order to regulate this piece, we can impose a bulk cutoff far outside the Dyson sphere; or we could consider the mutual information between $R$ and $\bar R-o$, the complement of $R$ with a small gap $o$ between $R$ and $\bar R$ removed, $I \equiv S_R+S_{\bar R-o} -S_{R\bar R-o}$.

Before the Page time, $\gamma(t)$ is similar to the RT surface expected for $R$ in the vacuum (Fig.~\ref{fig-localdyson}a). $EW(R)$ includes the reservoir and nothing else of relevance. Therefore, $S_R(t)$ will increase, commensurate with the entropy of the Hawking radiation that has arrived in the reservoir.\footnote{Gravitational backreaction from the changing mass of the reservoir could alter the area of $\gamma(t)$. We prevent this by initially filling the reservoir with unentangled ballast particles that are moved to distant regions on the Dyson sphere as the radiation is moved in.}
\begin{figure}%
    \centering
    \includegraphics[width=.6\textwidth]{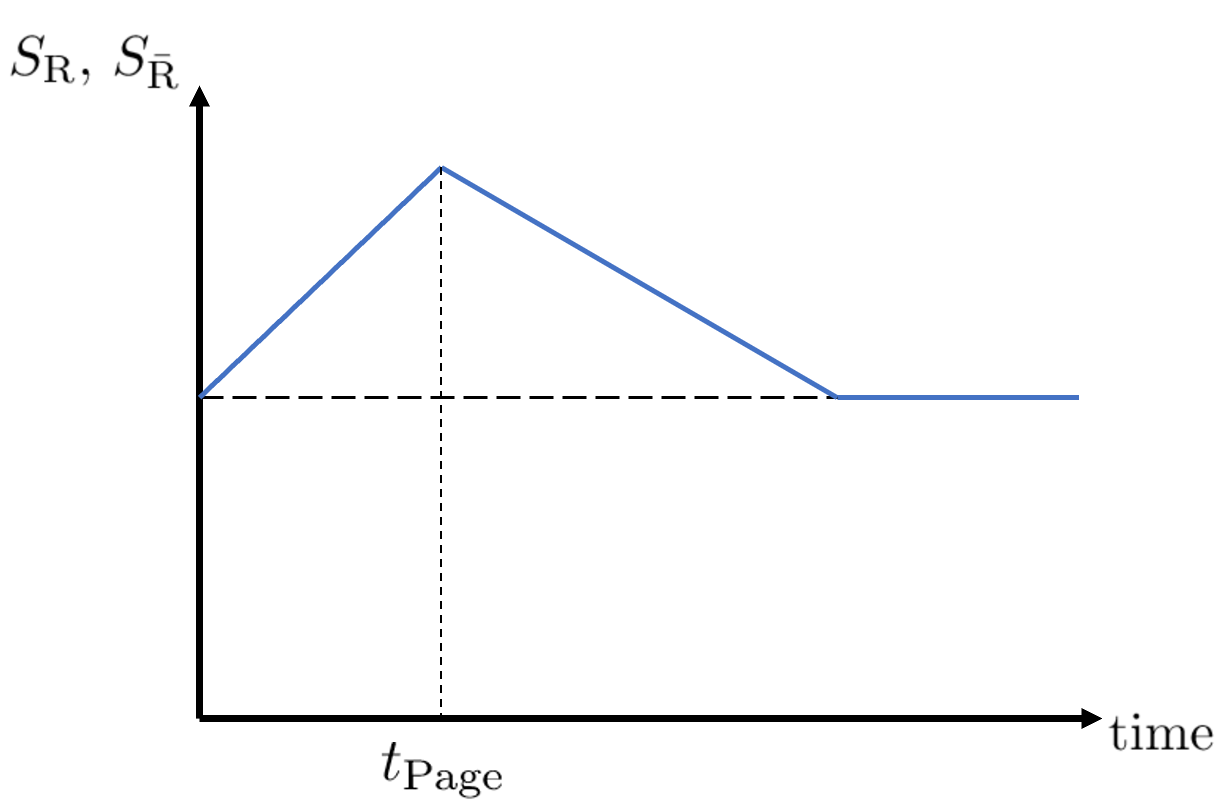}
    \caption{Up to a constant contribution from vacuum entanglement between $R$ and $\bar R$, the entropy of the two complementary boundary regions follows a Page curve. From the boundary point of view, this is because a system is slowly transferred from $\bar R$ to $R$. The RT prescription reproduces this curve from a bulk geometry obtained by semiclassical bulk evolution. However, this bulk dual is again inconsistent with the extrapolate dictionary (see Sec.~\ref{sec-interpretations}).}%
    \label{fig-pagecurve}%
\end{figure}

After the Page time, $\gamma(t)$ will have a second component, namely the new quantum extremal surface discovered in Refs.~\cite{Pen19,AEMM} (Fig.~\ref{fig-localdyson}b). This configuration is favored because inclusion of the interior Hawking partners in $EW(R)$ lowers its generalized entropy compared to the single-component quantum extremal surface anchored on $\partial R$. In this configuration, the bulk entropy of the Hawking radiation in the reservoir does not contribute to $S_R$ because its purification (the interior) is also in $EW(R)$. Hence the only dynamically relevant contribution comes from the area of the new quantum extremal surface component, i.e., the horizon area. We obtain the Page curve (Fig.~\ref{fig-pagecurve}) for the reservoir.

Though we have already argued that $S(\bar R)=S(R)$, it is instructive to verify directly that the Page curve results for $S(\bar R)$. Before the Page time $EW(\bar R)$ contains the black hole interior, but not the exterior Hawking radiation that has been absorbed into the reservoir. Hence the bulk matter entropy in $EW(\bar R)$ increases. After the Page time, $EW(\bar R)$ contains only the black hole exterior but not the reservoir, so there is neglible matter contribution. The time-dependent component of the RT surface is at the black hole horizon and so shrinks to zero at the required rate.

\section{Some Interpretations and Their Challenges}
\label{sec-interpretations}

In the previous section, we verified that a semiclassical bulk calculation, combined with the quantum-corrected RT prescription, yields CFT entropies consistent with unitary evolution~\cite{Pen19,AEMM}. In this section, we discuss a number of possible interpretations of this striking result. We argue that the semiclassically evolved bulk state is inconsistent with other aspects of the AdS/CFT correspondence.

\subsection{Bulk Information Loss vs.\ Extrapolate Dictionary}
\label{sec-infoloss}

We first discuss the most straightforward interpretation of the above results and, by extension, of Refs.~\cite{Pen19, AEMM}. We do not intend to ascribe this interpretation to these or other authors. Moreover, we will ultimately reject it. We consider it here because it is too obvious to ignore and so deserves discussion.

Let us take every aspect of the above analysis literally, at face value. That is, the boundary state evolves unitarily, from a pure state to a pure state. The global bulk state also evolves unitarily through the semiclassical equations of motion. But in the bulk state, the Hawking radiation (and thus the Dyson sphere) is entangled with the black hole interior. The reduced state of the Dyson sphere alone is therefore mixed, in accordance with Hawking's original prediction of information loss. 

The bulk state becomes pure only if one includes the black hole interior, which the RT prescription does automatically after the Page time. But an asymptotic bulk observer cannot access this region. The resulting picture is precisely that advocated by Unruh and Wald~\cite{UnrWal17}: information loss to the bulk observer, even though the boundary theory is unitary. 

Bulk information loss should simply be called information loss. When we carry out a scattering experiment, we are bulk observers. The relevant question is whether careful examination of the Hawking radiation allows us to reconstruct the initial state. This would not be the case if the bulk state at late times is truly the one determined by Hawking's calculation, as we assume here.

% This interpretation gains in plausibility through the insight of~\cite{Pen19,AEMM} that the RT proposal reproduces the boundary entropy expected from unitarity, despite the information loss in the bulk. 
The manifest unitarity of the CFT is usually viewed as a significant argument that information must be returned to a distant bulk observer when a black hole forms and evaporates. But here the RT prescription appears to render boundary unitarity perfectly consistent with bulk information loss.

The fact that the boundary state is pure does lead to a contradiction, however: not with RT, but with the standard AdS/CFT ``extrapolate'' dictionary. We now explain this contradiction.

For simplicity, let us first consider a time after the black hole has fully evaporated and all of the Hawking radiation has been absorbed into the Dyson sphere. The gravitational backreaction of the Dyson sphere can be made arbitrarily small.\footnote{This will be true even though the number of operators required to measure the out-state is $O(N^2)$. Backreaction from such a large number of operators invalidates the $1/N$ expansion only if they are all contained in a region bounded by area of comparible magnitude in Planck units. The Dyson sphere is much larger and the required operators can be arbitrarily dilute in space.} The bulk geometry will be classical to arbitrarily good approximation. Therefore, the boundary state can be computed from the bulk state using, for example, the methods of~\cite{HamKab05,HamKab06,KabLif11} (see also~\cite{BanDou98,HarSta11}).

In this standard dictionary, bulk operators that approach the boundary become boundary operators. The bulk density operator for the Dyson sphere is a mixture of pure states each of which can be created from the bulk vacuum using creation operators of wavepackets that are close to the boundary. Therefore the boundary state will be mixed, with the same entropy as the Dyson sphere.

Since the black hole is gone, the energy of the CFT state is fully accounted for by the energy of the Dyson sphere in the bulk. The CFT lives on a compact sphere, so any additional excitations would have a finite energy cost. Therefore there are no other, more diffuse CFT degrees of freedom available to purify the (arbitrarily large) entropy of the CFT excitations dual to the Dyson sphere.

Thus, the extrapolate dictionary demands that the CFT state is mixed at late times. This contradicts unitary evolution of the CFT on the boundary. Therefore, taken at face value, the analysis of the previous section (and hence of Refs.~\cite{Pen19,AEMM}) is not consistent with the established bulk-boundary dictionary.

This contradiction arises not only after the black hole has fully evaporated, but at all times after the Page time. In fact, the post-Page bulk state, computed semiclassically, does not correspond to {\em any\/} pure CFT state. Regardless of whether the bulk {\em evolution\/} actually proceeds as in Hawking's semiclassical calculation, one can consider the post-Page semiclassical state by formulating it as an {\em initial condition\/} in the bulk; and we claim that this bulk state has no boundary dual.

Again, this follows directly by applying the extrapolate dictionary to the Dyson sphere. Since the Dyson sphere is in a mixed state, the resulting boundary state must be in a mixed state. Any contributions from the remaining black hole cannot help. The entropy of the Dyson sphere is unbounded from above, at fixed mass of the remaining black hole. The bulk and boundary energy must agree, and the boundary energy is largely accounted for by the Dyson sphere, except for the finite mass of the black hole. Because the CFT has a finite number of states at any energy, this finite remainder is insufficient to purify the arbitrarily large entropy of the excitations dual to the Dyson sphere.

While it is only tangential to our argument, it is worth noting that this problem first arises at the Page time. The CFT degrees of freedom describing the black hole and those dual to the Dyson sphere approximately factorize, because the Dyson sphere is much closer to the boundary than the black hole horizon. By the extrapolate dictionary, the Dyson sphere corresponds to CFT operators localized to much less than the thermal wavelength associated with the black hole. Then by the known relation between energy and entropy of the CFT states dual to black holes, the number of these states becomes too small to purify the Dyson sphere at the Page time.
\begin{figure}%
    \centering
    \includegraphics[width=.5\textwidth]{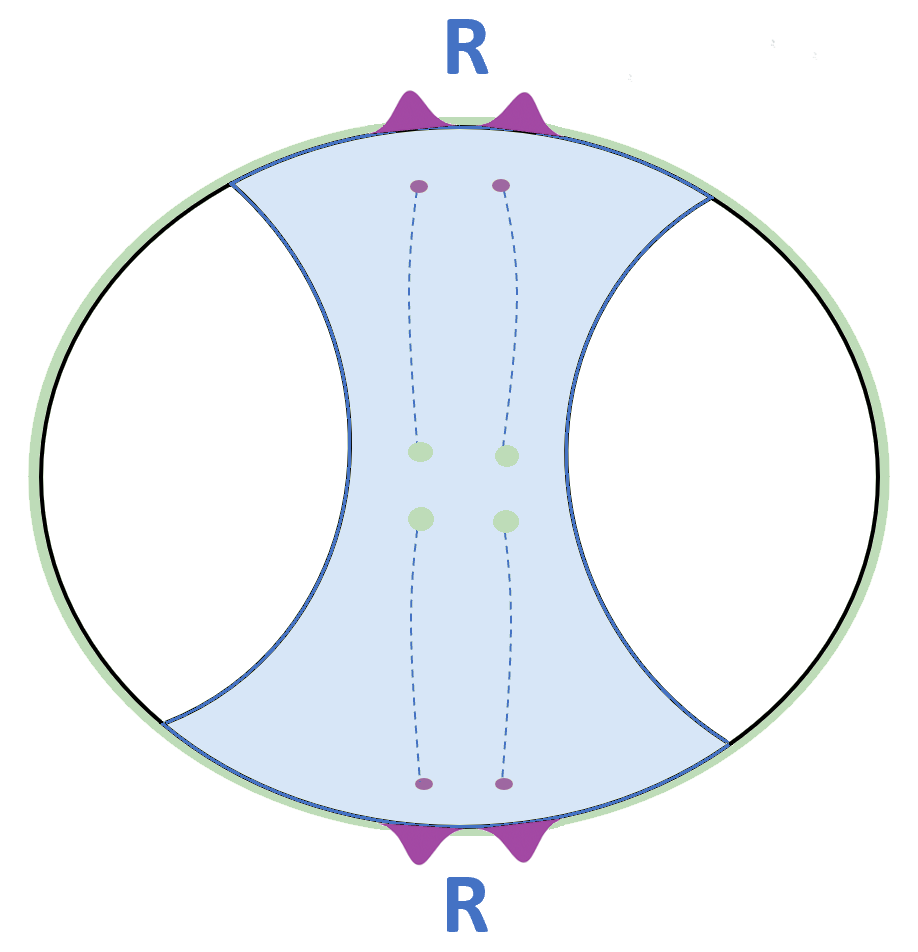}
    \caption{The entanglement wedge of a boundary region $R$ consisting of two components. Near-boundary particles are purified by particles deep in the bulk. This is consistent with a low-entropy state of $R$, since the deep particles have an energetic imprint on the boundary. Hence dilute CFT degrees of freedom are available to purify the more localized excitations. By contrast, bulk excitations behind a black hole horizon leave no energetic imprint near infinity, so there need not be enough states available in the CFT to represent them.}%
    \label{fig-twointervals}%
\end{figure}

It is worth noting that this situation differs from an interesting case studied recently in Ref.~\cite{AkeLev19}, shown in Fig.~\ref{fig-twointervals}.  In this case, the reconstruction wedge of a boundary region $R$ exceeds the causal wedge because near-boundary quanta are purified by quanta in the center of AdS. There, too, the extrapolate dictionary can be applied to the near-boundary excitations, implying the existence of localized boundary excitations in a mixed state. However, in that case the bulk purification is not behind a horizon. It has an energetic imprint on the boundary, associated with dilute boundary excitations that purify the localized ones. By contrast, in our example above, the purification is behind a horizon, and its entropy greatly exceeds the available CFT energy.

To summarize, the setting of Refs.~\cite{Pen19,AEMM}, taken at face value, implies bulk information loss. However, the boundary state is pure. We argued that this is incompatible with the standard AdS/CFT dictionary.

\subsection{Ambiguity in the Asymptotic Bulk State vs.\ QFT}
\label{sec-firstcommandment}

We will now consider, and again reject, a different interpretation of Refs.~\cite{Pen19,AEMM}, in which the bulk state is {\em not\/} taken at face value as the state predicted by semiclassical evolution. Instead, we allow for the possibility that for some purposes, the Hawking radiation as a whole must be considered to be in a pure state, and that sufficiently careful experiments by the asymptotic observer would confirm this.

In this viewpoint, the semiclassically evolved state invoked in the previous sections should be ``used'' only for some purposes, particularly for any questions posed by infalling observers. From the global state, upon tracing over the interior of the black hole, one obtains a thermal state for the Hawking radiation. This is good enough for simple probes (low-point functions) of the Hawking radiation, which cannot distinguish between a typical pure state and a thermal state. Presumably, one should also use the semiclassical state when applying the RT prescription, since it gives the desired answer (as shown in Refs.~\cite{Pen19,AEMM} and Sec.~\ref{sec-dyson} above).

But when we ask about careful measurements of the Hawking radiation by a bulk observer, we should use the pure state of the Hawking radiation predicted by a unitary S-matrix. That state should also be used when we apply the extrapolate dictionary to the state of the Dyson sphere, in order to avoid the conflict with the extrapolate dictionary that would otherwise result (see Sec.~\ref{sec-infoloss}). It is not clear what this state looks like globally---indeed, there are arguments that it is incompatible with a smooth horizon~\cite{Haw76,AMPS}. We consider it here as the state only of the Hawking radiation, to be contrasted with the thermal state discussed in the previous paragraph.

Naively, this seems like nonsense: no system described by quantum mechanics can be simultaneously in two different, distinguishable states. But what one means by ``a system'' can be a subtle question if nonlocal effects are important. The bulk contains both gravity and quantum mechanics. In that setting, one expects locality to be emergent, not fundamental. Therefore, we should not dismiss outright the possibility that two different approximate local descriptions may be valid for the Hawking radiation, depending on the question asked.

Nevertheless, we will argue that any ambiguity concerning the state of the bulk Hawking radiation is incompatible with an assumption much weaker than bulk locality---so weak, indeed, that we will call it an assertion. We assert that {\em the
%V2
out-state of a scattering experiment must admit a complete description in terms of a quantum field theory state in the algebra of the isolated region, in the sense of QFT on a curved background spacetime. All detector responses, for any simple or complicated measurement, can be predicted from a unique (``pure''\footnote{Here
%V2
  ``pure'' means that information about a pure in-state is returned. Strictly, the state of any finite region is mixed due to the usual vacuum entanglement. Here we ignore the vacuum entanglement across the boundary of the asymptotic region that is being probed by the detectors.} or mixed) state of the system.}\footnote{For predictions with probabilistic outcomes, the usual rules of quantum mechanics apply: the experiment must be repeated many times to verify the prediction.}
% V2

We claim that any corrections to the detector response are at least power-law suppressed by the ratio of system size to curvature radius in the system region. The strength of gravity far from the system, or at some earlier time, is irrelevant. If this were not the case, quantum field theory would be an uncontrolled approximation. It would be impossible to predict when it works, or to understand why it has been confirmed by all experiments so far. (It is not clear to us whether our assertion conflicts with the island proposal of Ref.~\cite{AMMZ}.) 

Here, the isolated system we have in mind is the bulk Hawking radiation, and the relevant weakly gravitating region is a shell occupied by this radiation. We assert that the outcome of a bulk scattering experiment is fully described as a state in quantum field theory at late times. Effects from quantum gravity can be made negligible. The quantum state could be pure as demanded by a unitary S-matrix; or it could be mixed as predicted by semiclassical bulk evolution. But {\em it cannot be necessary to invoke two different states to describe different experiments that may be performed on the Hawking radiation alone.}

Let us discuss this in detail in asymptotically flat spacetime. A black hole is formed by collapse of a star in the pure quantum state $|\psi\rangle \langle\psi|$. The black hole will evaporate and disappear, leaving behind a cloud of Hawking radiation in some state $\rho$. The radiation cloud can become arbitrarily large and dilute, so that $\rho$ can be thought of as the out-state in the asymptotic sense of the S-matrix.

In practice, the out-state is measured at finite time by a finite detector. But interactions and gravitational backreaction can be made arbitrarily small in the late-time, large-distance limit. At the very least, we require that the radiation be measured at a time much greater than the evaporation time scale $\Delta t_{\rm evap}\sim R^3/l_P^2$, where $R$ is the initial black hole radius and $l_P$ is the Planck length. At this time, the Hawking cloud is a shell of thickness $\Delta t_{\rm evap}$ and much greater radius.

Even if the evaporation process is not complete, our assertion applies to the regions occupied by the Hawking radiation. Since $t_{\rm evap}\gg R$, most of the radiation is far from the black hole. Thus, for the case of unitary evaporation, our assertion applies to complicated decoding operations after the Page time that extract scrambled quantum information into physical qubits.

Lest our assertion be misinterpreted as a stronger statement than it is, we would like to add the following clarifications:
\begin{itemize}
\item Our assertion does not preclude black hole complementarity. It allows for the possibility that complicated degrees of freedom in the Hawking radiation admit another interpretation as local degrees of freedom in the black hole interior. We assert only that there exists a unique (pure or mixed) QFT state in the algebra associated to the distant region containing the Hawking radiation, from which every possible detector response can be predicted, including the response when complicated measurements are performed, {\em in the distant region}.
\item We do not, of course, claim that QFT must describe the entire scattering process. If gravity becomes strong in some regions at intermediate times, our assertion does not apply to those regions. In particular, locality may break down substantially near or inside of a black hole. If the late-time state of the Hawking radiation is pure, we do not claim that a bulk QFT can explain how this came about. We assert only that there must exist a QFT state that fully describes how detectors respond to the Hawking radiation {\em at late times}, in all simple or complicated experiments.
\item For many simple experiments, a coarse-grained state will lead to the same predictions as the actual fine-grained state of the system. Our assertion does not contradict this. We assert only that there exists a unique state that correctly predicts {\em all} detector responses in {\em any} experiment, simple or complicated. This is what is always meant by the state of the system in quantum mechanics.
\end{itemize}

We close with the following disclaimers:
\begin{itemize}  
\item Our assertion applies to the distant Hawking radiation in a real-world experiment. One can imagine devising toy models where quantum gravity effects are always important in the region occupied by the Hawking radiation, so that there is no accurate or unambiguous QFT description of the out-state at any time. This may alleviate the information paradox, but it would do so by changing the problem. From such a model, no reliable conclusions could be drawn for the study of the information paradox for a black hole experiment in a real laboratory, where the QFT out-state is unambiguous.
\item It is conceivable (though, to us, implausible) that the formation and evaporation of a black hole in AdS is fundamentally different from the same process in Minkowski space, even when a Dyson sphere or auxiliary systems are included to absorb the radiation. In that case the previous remark would apply: we would be unable to draw reliable conclusions about the real-world information paradox from AdS/CFT.
\end{itemize}

%We now apply this assertion to the setting of Sec.~\ref{sec-}. We conclude that either the AdS setting is useless for analyzing the real-world information paradox, or the Dyson sphere is fully described by a unique (pure or mixed) quantum state. Our assertion excludes the possibility that different quantum states must be invoked to describe different probes of the Dyson sphere. It follows that the Hawking radiation is truly and unambiguously mixed. But as we argued in Sec.~\ref{sec-}, this possibility is excluded by the standard AdS/CFT dictionary, since the boundary state is pure.

%To summarize, we argued here that the semiclassically evolved bulk state considered in Refs.~\cite{Pen19,AEMM} must be taken at face value at least for the Hawking radiation. Nonlocal effects cannot be large in distant regions where gravity is weak. the latter is unambiguously in a mixed state. We showed in the previous section that this implies a mixed state  leads  is both inconsistent with the AdS/CFT duality.

\subsection{Pure Out-State From Nonlocal Interactions of Asymptotic Detectors?}
\label{sec-dual}

Finally, we consider the possibility that the Hawking radiation unambiguously returns the information to the outside bulk observer (the Dyson sphere). In asymptotically flat spacetime, this is just the statement that the S-matrix is unitary. In the setting of Sec.~\ref{sec-dyson}, it is the statement that there is a unitary map
\begin{equation}
  |\psi\rangle_{\rm in} \otimes |0\rangle_{\rm Dyson}(t_{\rm early}) \to |\Psi\rangle_{\rm Dyson} (t_{\rm late})~,
  \label{eq-unitary}
\end{equation}
where $|\psi\rangle_{\rm in}$ is the in-state of the collapsing matter, $|0\rangle_{\rm Dyson}$ is the initial fiducial state of the Dyson sphere, and $|\Psi\rangle_{\rm Dyson}$ is the final state of the Dyson sphere after it has absorbed all of the Hawking radiation. All of the above states live in a weakly gravitating region and so can be interpreted as states in QFT on a curved background. All detector responses are determined by the standard rules of local QFT.

But how is this compatible with the bulk calculation in Sec.~\ref{sec-dyson}? There we applied the RT prescription to a semiclassically evolved state after the Page time. It was crucial in this analysis that the final state of the Dyson sphere was {\em not\/} determined by Eq.~(\ref{eq-unitary}). Rather, the Hawking radiation was explicitly entangled with the black hole interior, and by itself was thermal. The Dyson sphere was in a mixed state at the end. This was essential for the RT analysis to yield the results found in Sec.~\ref{sec-dyson}, consistent with boundary unitarity.

One could speculate as follows.\footnote{We do not claim originality for this interpretation; but we also do not intend to ascribe it to anyone.} The final state in Eq.~(\ref{eq-unitary}) is an ``effective'' description that results from integrating out nonlocal degrees of freedom in the black hole interior that are not accessed. This ``effective description'' is capable of describing all detector responses to any experiment that might be performed on the Hawking radiation, using the standard rules of local QFT in the asymptotic region. (Thus, this interpretation is not in conflict with the assertion of the previous subsection that there must exist a unique state like this.) However, there is another, dual description of such bulk experiments, in which the late time bulk state is taken to be the semiclassically evolved state of Sec.~\ref{sec-dyson}, and in which asymptotic detectors respond nonlocally to the black hole interior (or more generally, to the distant semiclassical ``islands'' picked out by the RT prescription in Refs.~\cite{Pen19,AEMM} and by Eq.~15 in Ref.~\cite{AMMZ}). In this alternate description, therefore, detectors in a weakly gravitating region do {\em not} respond to the assumed quantum state (the semiclassical state) as demanded by local quantum field theory.

It is not clear to us what advantage is gained by the second viewpoint. In the asymptotic region, a simpler theory is already available: in the state $|\Psi\rangle_{\rm Dyson}(t_{\rm late})$, the usual rules of QFT determine all detector responses. There is no reason to introduce a more convoluted description, in which the state is different and detectors respond in nonstandard ways. And at the black hole horizon, the apparent smoothness of the semiclassically evolved state does not guarantee that an infalling observer actually experiences no drama, because detector response is not determined by the standard rules of local QFT.

The validity of the first (pure-state) description for the Hawking radiation is not questioned in this interpretation. The AMPS argument then implies that a smooth horizon is inconsistent with the linearity of quantum mechanics at late times~\cite{AMPS,Bou12c,AMPSS}. For this argument it is sufficient that the standard description of a pure out-state is valid (along with other assumptions stated in Ref.~\cite{AMPS}). The existence of any dual descriptions is irrelevant. Getting rid of the firewall then requires significant nonlinearity through state-dependence~\cite{Bou13,MarPol13,Bou13a}. See Refs.~\cite{PapRaj13b, MalSus13} for interesting approaches. Such ideas can be considered independently of the present discussion. If they can be developed into a consistent theory, a standard local description of the asymptotic region appears to be sufficient.

\section{Ryu-Takayanagi For Unitary Evaporation}
\label{sec-unitary}

In Sec.~\ref{sec-dyson}, we applied the RT prescription to the semiclassical bulk state $\rho_{\rm Hawking}$. We found that it computes a boundary entropy consistent with unitary evolution. In Sec.~\ref{sec-interpretations}, however, we found that bulk loss of information is inconsistent with a pure boundary state for other reasons. We further argued that $\rho_{\rm Hawking}$ is unambiguous in the asymptotic region, and hence is incompatible with bulk evolution leading to a pure out-state $|\Psi_{\rm Dyson}\rangle (t_{\rm late})$. Assuming boundary unitarity, we are forced to consider bulk states in which the Dyson sphere at late times is in the state $|\Psi_{\rm Dyson}\rangle (t_{\rm late})$.

This does not tell us about the rest of the bulk yet. An interesting question is how a bulk state that reduces to $|\Psi_{\rm Dyson}\rangle (t_{\rm late})$ on the Dyson sphere could be ``completed'' to a global state in such a way that the RT proposal succeeds in computing the boundary entropy.
That is, we would like the RT proposal to return 0 for the entropy of the entire boundary, at all times; and in the refined setup of Sec.~\ref{sec-fancy}, we would like to obtain the Page curve for both boundary portions. How does this constrain the global bulk state?

Consider the Dyson sphere at the intermediate time $t$, after it has absorbed a fraction $\alpha$ of the total amount of Hawking radiation. For simplicity, we will neglect the irreversibility of the evaporation and the time-dependence of the temperature; neither will be important here. Before the Page time, $S_{\rm rad}=\alpha S_0$, where $S_0=A_0/4G\hbar$ is the initial entropy of the black hole; after the Page time, $S_{\rm rad}=(1-\alpha) S_0$.

Before the Page time, there is a consistent bulk spacetime with smooth horizon which satisfies all relevant constraints: the one that would be computed semiclassically. We are not claiming that this is exactly the correct state. But this state is consistent with $S_{\rm rad}=\alpha S_0$; and it is also consistent with $S_{CFT}=0$, via the RT prescription, since the dominant quantum extremal surface is the empty surface, and the interior Hawking partners purify the exterior ones in the Dyson sphere.

After the Page time, however, there does not appear to be a consistent bulk spacetime with smooth horizon. Bulk unitarity demands $S_{\rm rad}=(1-\alpha) S_0$. (This is also demanded by boundary unitarity combined with the extrapolate dictionary applied to the boundary subsystem dual to the Dyson sphere.) But the full boundary entropy vanishes, $S_{CFT}=0$. By Eq.~(\ref{eq-rt}), we have
\begin{equation}
  0 = \frac{A_{RT}}{4G\hbar}+ S_{\rm bulk}~.
  \label{eq-flm0}
\end{equation}
Any macroscopic, minimal quantum extremal surface homologous to the full boundary would contribute an area term and so would make this equation impossible to satisfy. However, there appears to be no such surface (regardless of the minimality condition). The existence of the nontrivial quantum extremal surface of Ref.~\cite{Pen19,AEMM} relied on the entanglement properties of the semiclassically evolved state, specifically on the entanglement of inside and outside Hawking partners. Now that we assume that the bulk out-state is pure, this entanglement pattern is inconsistent with the fact that outgoing Hawking particles can be almost purified by the early Hawking radiation after the Page time~\cite{AMPS}. 

With $A_{RT}=0$, Eq.~(\ref{eq-flm0}) still requires that $S_{\rm bulk}=0$. The Hawking radiation absorbed in the Dyson sphere is a subsystem of the bulk, with entropy $S_{\rm rad}=(1-\alpha) S_0$. In order to have $S_{\rm bulk}=0$, we must invoke another bulk system $F$ within the entanglement wedge which purifies the Dyson sphere. This system must therefore have entropy $S_F=(1-\alpha) S_0$,

We are not able to identify such a system consistent with smoothness of the horizon and with at least approximate validity of semiclassical evolution laws in regions of low curvature. For example, the interior Hawking partners of the future Hawking radiation would have the right amount of entropy. But they would not be available for our present task even if the horizon were smooth. This is because they would then be purified by the future outside radiation, not by the radiation that is already in the Dyson sphere.

Indeed, the mere presence of information outside of the black hole is inconsistent with a smooth bulk and approximate validity of the semiclassical equations in low-curvature regions. This is because the latter would require the presence of the star in the deep interior of the black hole. Unitarity of the out-state after the Page time would then imply that some of the quantum information in the star has been duplicated. In the context of complementarity~\cite{SusTho93}, this was not a problem, because no observer could see both copies~\cite{SusTho93b}. However, in the context of the RT proposal it {\em is\/} a problem, because the size of the entanglement wedge is unconstrained by the limitations of causal observers.  

If the RT proposal can be applied at all after the Page time, then the above considerations lead to a dramatic conclusion. Semiclassical evolution must break down substantially in some low-curvature region, i.e., it must fail completely as an approximate description. A firewall at the horizon~\cite{AMPS} would be a special case of such a failure. Assuming that its entropy is given by the Bekenstein-Hawking entropy of the horizon, this would naturally provide a system with entropy $S_F=(1-\alpha) S_0$ that could purify the Dyson sphere.

It is possible that the breakdown of semiclassical gravity occurs in some other low-curvature region, for example deeper inside or even outside of the black hole. Since it is this failure that makes firewalls so unpalatable, these alternatives offer little comfort. Moreover, their entropy would not be related to the Bekenstein-Hawking entropy.

We conclude that the simplest, most conservative choice of purification of the Dyson sphere consistent with both the RT proposal and the extrapolate dictionary is to end spacetime at the horizon (a firewall). This interior boundary should be treated not as an extremal surface, but as an object that together with the Dyson sphere is in the pure state demanded by unitarity.

\section{Discussion}
\label{sec-discussion}

In Sec.~\ref{sec-dyson}, we reproduced the results of Refs.~\cite{Pen19,AEMM} in a setting that eliminates a key assumption about entanglement wedge complementarity. By applying the quantum-corrected RT prescription to a bulk state obtained from semiclassical evolution, we found that appropriate boundary regions obey the Page curve, consistent with unitary boundary evolution.

This result is highly suggestive of a resolution of the information paradox: the vanishing boundary entropy suggests that unitarity is maintained; yet the semiclassical bulk state has a smooth horizon. In Sec.~\ref{sec-interpretations}, however, we noted that unitarity of the {\em bulk\/} S-matrix is manifestly violated in the setting of Refs.~\cite{Pen19,AEMM}; and we found that restoring the ability to recover information to an asymptotic bulk observer (not just a boundary observer) appears to require significant modifications to the S-matrix framework in which the information paradox is normally posed. Let us discuss this in more detail.

The most straightforward interpretation of Refs.~\cite{Pen19,AEMM} is that information is lost to a bulk observer (even when careful experiments are performed) but retained in the boundary theory (Sec.~\ref{sec-infoloss}). We stress that this would be tantamount to information loss in real-world scattering experiments. We then considered two interpretations that would allow a careful bulk observer to recover the information. We found that both involve significant new physics in asymptotic regions.

If the semiclassical out-state is to be used for some questions and the unitary out-state for others, then there is no unique quantum state describing all experiments that can be performed in the asymptotic region, and hence there is no S-matrix in any strict sense (Sec.~\ref{sec-firstcommandment}). If the semiclassical quantum state is viewed as fundamental (Sec.~\ref{sec-dual}), then an ``effective'' unitary out-state must arise from nonlocal couplings of asymptotic detectors to the black hole interior in the state $\rho_{\rm Hawking}$. This means that the validity of QFT in arbitrarily distant weakly gravitating regions depends on the history of the quantum state, which is a significant departure from standard physics and renders QFT essentially an uncontrolled approximation. Moreover, detectors respond nonlocally in the semiclassical state. Thus, we cannot conclude without a more complete description that the horizon would actually be smooth (since this conclusion assumes a standard local detector response).

At first, the above implications of Refs.~\cite{Pen19,AEMM} are reminiscent of other proposals that identify the black hole interior with the distant radiation in a state-dependent way, such as ER=EPR~\cite{MalSus13} and the Papadodimas-Raju construction of interior operators~\cite{PapRaj12,PapRaj13b}. However, these approaches ascribed the new physics to the black hole region, not to the asymptotic region. The unitary out-state is unambiguous and fundamental.

The appearance of new physics near the horizon of an arbitrarily large hole is already troubling, of course. But in ER=EPR and Papadodimas-Raju, at least we can define a regime of validity for standard physics: the presence of a sufficiently old black hole furnishes a quasilocal criterion for the breakdown of standard local effective field theory.

By contrast, in seeking to interpret Refs.~\cite{Pen19,AEMM} as a resolution of the information paradox, we are forced to change physics in asymptotic regions. When presented with a dilute cloud of radiation and a quantum state for it, we would never know ahead of time whether our detectors will respond according to the standard rules of QFT.

If we are reluctant to accept such modifications, what should we make of the apparent success of the computation in Refs.~\cite{Pen19,AEMM}? These are nontrivial, highly intriguing results, and they need to be understood. The main goal of our work was to point out that they cannot be straightforwardly read as a resolution of the information paradox without introducing significant new physics in asymptotic regions.

We conclude with a speculation about an alternative interpretation. In this paper, we have assumed at all times that the boundary evolution is unitary, as it would be if the boundary theory is a particular unitary field theory. However, there is significant evidence that JT gravity, in which the most explicit results have been obtained~\cite{AEMM}, is dual not to a unique unitary theory but to an ensemble of theories~\cite{SaaShe18,SaaShe19}.

Let us suppose that the theories in the ensemble have naturally identifiable in- and out-states, but they differ in the details of the interactions. Boundary evolution by the {\em ensemble\/} of theories would not be unitary: different members of the ensemble would evolve the same in-state to different out-states. So the ensemble of out-states would be mixed. This would be consistent with obtaining a thermal out-state in the bulk. Yet, each member of the theory ensemble is unitary, so the ensemble average of the late-time entropies would vanish. This would be consistent with the computation of the entropy by RT.\footnote{The RT result can be explicitly verified by a direct calculation of the Renyi entropies through a replica path integral. We thank A.~Almheiri, T.~Hartman, J.~Maldacena, H.~Maxfield, G.~Penington and D.~Stanford for discussions and presentations of ongoing projects which demonstrate this.}

It is not obvious to us that this is the correct interpretation. But we find it interesting that it would remove the key difficulties with reconciling bulk information loss and an (incorrectly assumed) boundary unitarity. However, it would mean that the firewall paradox remains. A real scattering experiment in which information is returned would be described by only one member of the ensemble (the ``correct'' theory), and so would not correspond to semiclassical bulk evolution.

\noindent {\bf Acknowledgements.}~We would like to thank C.~Akers, A.~Almheiri, V.~Chandra\-sekaran, R.~Emparan, N.~Engelhardt, T.~Hartman, A.~Levine, J.~Maldacena, H.~Maxfield, G.~Penington, A.~Shah\-bazi-Moghaddam, R.~Myers, P.~Rath, and D.~Stanford for very helpful discussions and comments.  We are grateful to T.~Banks, E.~Flanagan and B.~Freivogel for their comments on an earlier version of this manuscript. This work was supported in part by the Berkeley Center for Theoretical Physics; by the Department of Energy, Office of Science, Office of High Energy Physics under QuantISED Award DE-SC0019380 and under contract DE-AC02-05CH11231; and by the National Science Foundation under grant PHY1820912. MT acknowledges financial support from the innovation program under ERC Advanced Grant GravBHs-692951. 

\bibliographystyle{utcaps}
\bibliography{all}

\end{document}